\documentclass[final,5p,times,twocolumn]{elsarticle}
\usepackage{graphicx,color,epsfig,rotate,amssymb}
\usepackage{color,soul}
\definecolor{darkblue}{rgb}{0,0,0.5}
\setulcolor{darkblue}

\journal{Solid State Communications}

\begin{document}

\begin{frontmatter}

\title{The Role of Intra- and Inter-site exchange correlations in the Extended Falicov-Kimball Model on a Triangular Lattice}

\author[iitr]{Sant Kumar}
\author[iisc]{Umesh K. Yadav}
\author[iitr]{T. Maitra}
\author[iitr]{Ishwar Singh}
\address[iitr]{Department of Physics, Indian Institute of Technology Roorkee, Roorkee-247667, Uttarakhand, India}
\address[iisc]{Department of Physics, Centre for Condensed Matter Theory, Indian Institute of Science, Bangalore-560012, India}

\begin{abstract}
Ground state magnetic properties of the spin-dependent Falicov-Kimball model (FKM) are studied by incorporating the intrasite exchange correlation J (between itinerant $d$- and localized $f$- electrons) and intersite (superexchange) correlation $J_{se}$ (between localized $f$- electrons) on a triangular lattice for two different fillings. Numerical diagonalization and Monte-Carlo techniques are used to determine the ground state magnetic properties. Transitions from antiferromagnetic to ferromagnetic and again to re-entrant antiferromagnetic phase is observed in a wide range of parameter space. The magnetic moments of $d$- and $f$- electrons are observed to depend strongly on the value of $J$, $J_{se}$ and also on the total number of $d$- electrons ($N_d$).
\end{abstract}


\end{frontmatter}

\section{Introduction}

The problem of inhomogeneous charge and magnetic ordering in strongly correlated electron systems (SCES) is one of the most intensively studied problems of the contemporary condensed matter physics. The motivation behind these studies is the inhomogeneous charge ordering (e.g., striped phases) which has been experimentally observed in many rare-earth and transition-metal compounds like $La_{1.6}Nd_{0.4}Sr_{x}CuO_{4}$, $YBa_{2}Cu_{3}O_{6+x}$, $Bi_{2}Sr_{2}Cu_{2}O_{8+x}$~\cite{tran,mook}. Some of these compounds also exhibit high temperature superconductivity (HTSC). Theoretical studies on these materials proposed that these SCES have a natural tendency toward the phase separation~\cite{loren,leman}.

A class of SCES like cobaltates~\cite{qian06,tera97,tekada03}, $GdI_{2}$~\cite{tara08} and its doped variant $GdI_{2}H_{x}$~\cite{tulika06}, $NaTiO_{2}$~\cite{clarke98,pen97,khom05}, MgV$_{2}$O$_{4}$~\cite{rmn13} etc. have attracted great interest recently as they exhibit a number of remarkable cooperative phenomena such as valence and metal-insulator transition, charge, orbital and spin/magnetic order, excitonic instability and possible non-fermi liquid states~\cite{tara08}. These are layered triangular lattice systems and are characterized by the presence of localized (denoted by $f$-) and itinerant (denoted by $d$-) electrons. The geometrical frustration from the underying triangular lattice coupled with strong quantum fluctuations give rise to a huge degeneracy at low temperatures resulting in competing ground states close by in energy. Therefore, for these systems one would expect a fairly complex ground state magnetic phase diagram and the presence of soft local modes strongly coupled with the itinerant electrons.

It has recently been proposed that these systems may very well be described by different variants of the two-dimensional Falicov-Kimball model (FKM)~\cite{tara08,tulika06} on a triangular lattice. The FKM was introduced to study the metal-insulator transitions in the rare-earth and transition-metal compounds~\cite{fkm69,fkm70}. The model has also been used to describe a variety of many-body phenomenon such as tendency of formation of charge and spin density wave, mixed valence, electronic ferroelectricity and crystallization in binary alloys~\cite{lemanski05,farkov02}.

Many experimental results show that a charge order generally occurs with a spin/magnetic order. Therefore, the FKM on a triangular lattice has been studied recently including a spin-dependent on-site interaction between $f$- and $d$- electrons with a local Coulomb interaction between $f$- electrons. Several interesting ground state phases namely long range Ne\`el order, ferromagnetism or a mixture of both have been reported~\cite{umesh6,sant}.

It has been realized later that even though including the local spin-dependent interactions to the FKM on triangular lattice gives many interesting phases, in fact there are other important interactions e.g., super-exchange interaction ($J_{se}$) (interaction between $f$- electrons occupying nearest neighboring sites) between $f$- electrons, which gives rise to many other interesting phases relevant for real materials such as $GdI_{2}$,   $NaTiO_{2}$ etc.

Therefore, we have generalized the FKM Hamiltonian to include super-exchange interaction $J_{se}$ between $f$- electrons and Hamiltonian is given as,

\begin{eqnarray}
H  =-\,\sum\limits_{\langle ij\rangle\sigma}(t_{ij}+\mu\delta_{ij})d^{\dagger}_{i\sigma}d_{j\sigma}
+\,(U-J)\sum\limits_{i\sigma}f^{\dagger}_{i\sigma}f_{i\sigma}d^{\dagger}_{i\sigma}d_{i\sigma}
\nonumber \\
+\,U\sum_{i\sigma}f^{\dagger}_{i,-\sigma}f_{i,-\sigma}d^{\dagger}_{i\sigma}d_{i\sigma}
+\,J_{se}\sum\limits_{\langle ij \rangle \sigma} (-\,f_{i\sigma}^{\dagger} f_{i\sigma} f_{j,-\sigma}^{\dagger} f_{j,-\sigma}
\nonumber \\
+ \,f_{i\sigma}^{\dagger} f_{i\sigma} f_{j\sigma}^{\dagger} f_{j\sigma})
+\,U_{f}\sum\limits_{i\sigma}f^{\dagger}_{i\sigma}f_{i\sigma}f^{\dagger}_{i,-\sigma}f_{i,-\sigma}
+\,E_{f}\sum\limits_{i\sigma}f^{\dagger}_{i\sigma}f_{i\sigma}
\end{eqnarray}
\noindent here $\langle ij\rangle$ denotes the nearest neighboring ($NN$) lattice sites $i$ and $j$. The $d^{\dagger}_{i\sigma},  d_{i\sigma}\,(f^{\dagger}_{i\sigma},f_{i\sigma})$ are, respectively, the creation and annihilation operators for $d$- ($f$-) electrons with spin $\sigma=\{\uparrow,\downarrow\}$ at the site $i$. First term is the band energy of the $d$- electrons. Here $\mu$ is chemical potential. The hopping parameter $t_{\langle ij\rangle} = t$ for $NN$ hopping and zero otherwise. The interaction between $d$- electrons is neglected.  Second term is on-site interaction between $d$- and $f$- electrons of same spins with coupling strength ($U - J$) (where $U$ is the usual spin-independent Coulomb interaction and $J$ is the exchange interaction. Inclusion of the exchange term enables us to study the magnetic structure of $f$- electrons and band magnetism of $d$- electrons. Third term is the on-site interaction $U$ between $d$- and $f$- electrons of opposite spins. Fourth term is the super-exchange interaction between localized electrons occupying nearest neighboring sites. This interaction favors anti-ferromagnetic arrangement of $f$- electrons over ferromagnetic arrangement of $f$- electrons. Fifth term is on-site Coulomb repulsion $U_f$ between opposite spins of $f$- electrons. The last term is the dispersionless energy level $E_f$ of $f$- electrons.

\section{Methodology}

Hamiltonian $H$ (Eq.$1$), preserve states of the $f$- electrons, i.e. the $d$- electrons traveling through the lattice change neither occupation numbers nor spins of the $f$- electrons. Therefore, the local $f$- elctron occupation number $\hat{n}_{fi\sigma}=f_{i\sigma}^{\dagger}f_{i\sigma}$ is invariant and $\big[\hat{n}_{fi\sigma},H\big]=0$ for all $i$ and $\sigma$. This shows that $\omega_{i\sigma}=f_{i\sigma}^{\dagger}f_{i\sigma}$ is a good quantum number taking values only $1$ or $0$ according to whether the site $i$ is occupied or unoccupied by $f$- electron of spin $\sigma$, respectively. Following the local conservation of $f$- electron occupation, $H$ can be rewritten as,
\begin{eqnarray}
H&=&\sum\limits_{\langle ij \rangle \sigma}\, h_{ij}(\{\omega_{\sigma}\})\,d_{i\sigma}^{\dagger}d_{j\sigma}
+\,J_{se}\sum\limits_{\langle ij \rangle \sigma}{\{-\,\omega_{i\sigma}\omega_{j,-\sigma}}
\nonumber \\&&
+\,{\omega_{i\sigma}\omega_{j\sigma}\}}
+\,U_{f}\sum\limits_{i\sigma}{\omega_{i\sigma}\omega_{i,-\sigma}}
+\,E_{f}\sum\limits_{i\sigma}\,{\omega_{i\sigma}}
\end{eqnarray}
\noindent where $h_{ij}(\{\omega_{\sigma}\})=\big[-t_{ij}+\{(U-J)\omega_{i\sigma}+U\omega_{i,-\sigma}-\mu\}\delta_{ij}\big]$ and $\{\omega_{\sigma}\}$ is a chosen configuration of $f$- electrons of spin $\sigma$.


We set the scale of energy with $t_{\langle ij \rangle} = 1$. The value of $\mu$ is chosen such that the filling is ${\frac{(N_{f}~ + ~N_{d})}{4N}}$ (e.g. $N_{f} + N_{d} = N$ is one-fourth case and $N_{f} + N_{d} = 2N$ is half-filled case etc.), where $N_{f} = (N_{f_{\uparrow}}+N_{f_{\downarrow}})$, $N_{d} = (N_{d_{\uparrow}} + N_{d_{\downarrow}})$ and $N$ are the total number of $f-$ electrons, $d-$ electrons and sites respectively. For a lattice of $N$ sites the $H(\{\omega_{\sigma}\})$ (given in Eq.2) is a $2N\times 2N$ matrix for a fixed configuration $\{\omega_{\sigma}\}$. For one particular value of $N_f(= N_{f_{\uparrow}} + N_{f_{\downarrow}})$, we choose values of $N_{f_{\uparrow}}$ and $N_{f_{\downarrow}}$ and their configuration $\{\omega_{\uparrow}\} = \{{\omega_{1\uparrow}, \omega_{2\uparrow},\ldots, \omega_{N\uparrow}}\}$ and $\{\omega_{\downarrow}\} = \{{\omega_{1\downarrow}, \omega_{2\downarrow},\ldots, \omega_{N\downarrow}}\}$. Choosing the parameters $U$, $J$ and $J_{se}$, the eigenvalues $\lambda_{i\sigma}$($i = 1 \ldots N$) of $h(\{\omega_{\sigma}\})$ are calculated using the numerical diagonalization technique on the triangular lattice of finite size $N(=L^{2}, L = 12)$ with periodic boundary conditions (PBC).

The partition function of the system is written as,
\begin{equation}
\it{Z}=\,\sum\limits_{\{\omega_{\sigma}\}}\,Tr\,\left(e^{-\beta H(\{\omega_{\sigma}\})}\right)
\end{equation}
\noindent where the trace is taken over the $d$- electrons, $\beta=1/k_{B}T$. The trace is calculated from the eigenvalues $\lambda_{i\sigma}$ of the matrix $h(\{\omega_{\sigma}\})$ (first term in Eq.2).
\noindent The ground state total internal energy $E(\{\omega_{\sigma}\})$ is calculated as,
\begin{eqnarray}
E(\{\omega_{\sigma}\})=\sum\limits_{i\sigma}^{N_{d}}\lambda_{i\sigma}(\{\omega_{\sigma}\})
+ J_{se}\sum\limits_{\langle ij \rangle \sigma}{\{-\omega_{i\sigma}\omega_{j,-\sigma}}
+{\omega_{i\sigma}\omega_{j\sigma}\}}
\nonumber \\
+ U_{f}\sum\limits_{i\sigma}\omega_{i\sigma}\omega_{i,-\sigma}
+ E_{f}\sum\limits_{i\sigma}\omega_{i\sigma}
\end{eqnarray}

Our aim is to find the unique ground state configuration (state with minimum total internal energy $E(\{\omega_{\sigma}\}$)) of $f$- electrons out of exponentially large possible configurations for a chosen $N_{f}$. In order to achieve this goal, we have used classical Monte Carlo simulation algorithm by annealing the static classical variables $\{\omega_{\sigma}\}$ ramping the temperature down from a high value to a very low value. Details of the method can be found in our earlier papers~\cite{sant1,umesh1,umesh2,umesh3,umesh4,umesh5}.

\section{Results and discussion}

\begin{figure*}
{
\begin{center}
\includegraphics[trim=0.5mm 0.5mm 0.5mm 0.1mm,clip,width=7.0cm,height=4.8cm]{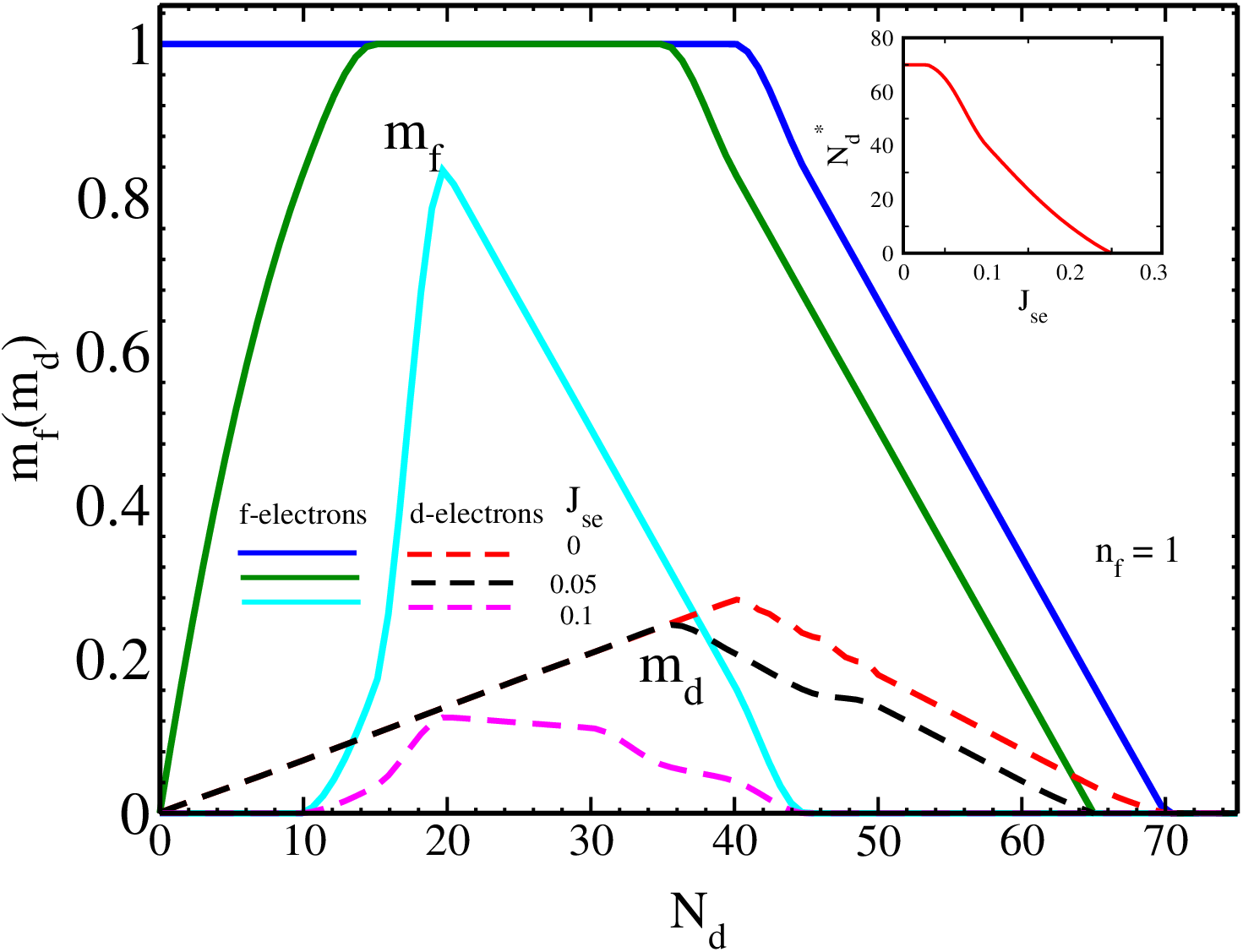}~~~~\includegraphics[trim=0.5mm 0.5mm 0.5mm 0.05cm,clip,width=8.9cm,height=4.5cm]{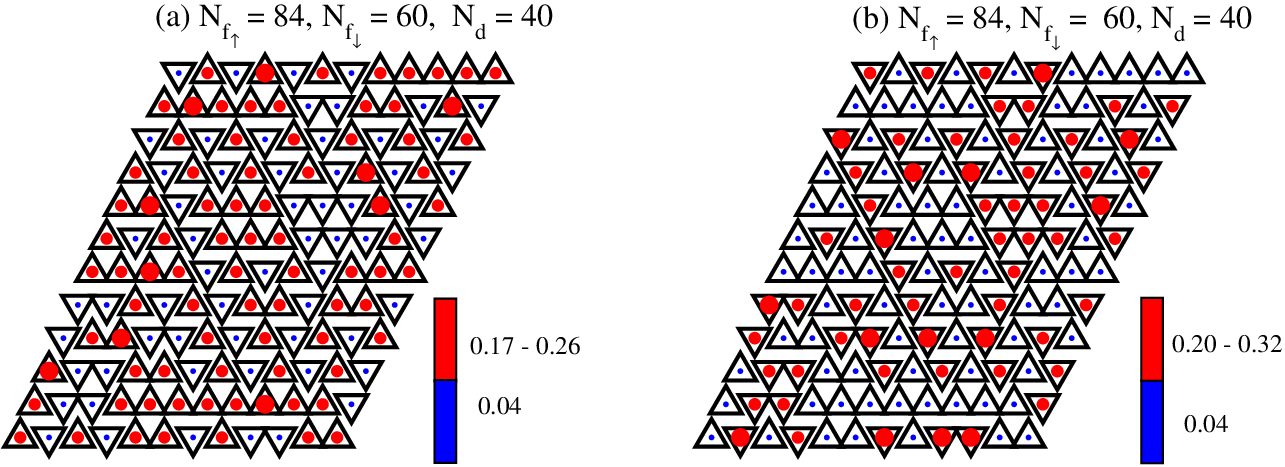}
{{\bf (i)}~~~~~~~~~~~~~~~~~~~~~~~~~~~~~~~~~~~~~~~~~~~~~~~~~~~~~~~~~~~~~~~~~~~~~~~~~~~~~~~~~~~~~~~~~~~~~~~~~~~~~~~~~{\bf (ii)}}
\end{center}
}
\caption{(Color online) {\bf (i)} Variation of magnetic moment of $d$- and $f$- electrons with number of $d$- electrons $N_d$ for $n_{f}=1$, $U = 5$, $J = 5$, $U_{f} = 10$ and for different values of $J_{se}$. Magnetic moment of $d$- and $f$- electrons are shown by dash and solid lines respectively. Variation of $N_{d}^{\ast}$ with $J_{se}$ is shown in the inset.} {\bf (ii)} (a) Up-spin and (b) down-spin $d$- electron densities are shown on each site for $J_{se} = 0.10$, $U = 5$, $U_{f} = 10$, $J = 5$, $n_{f} = 1$ and $N_{d} = 40$. The color coding and radii of the circles indicate the $d$- electron density profile. Triangle-up and triangle-down correspond to the sites occupied by up-spin and down-spin $f$- electrons respectively.
\end{figure*}

\begin{figure*}
{
\begin{center}
\includegraphics[trim=0.5mm 0.5mm 0.5mm 0.1mm,clip,width=8.9cm,height=6.0cm]{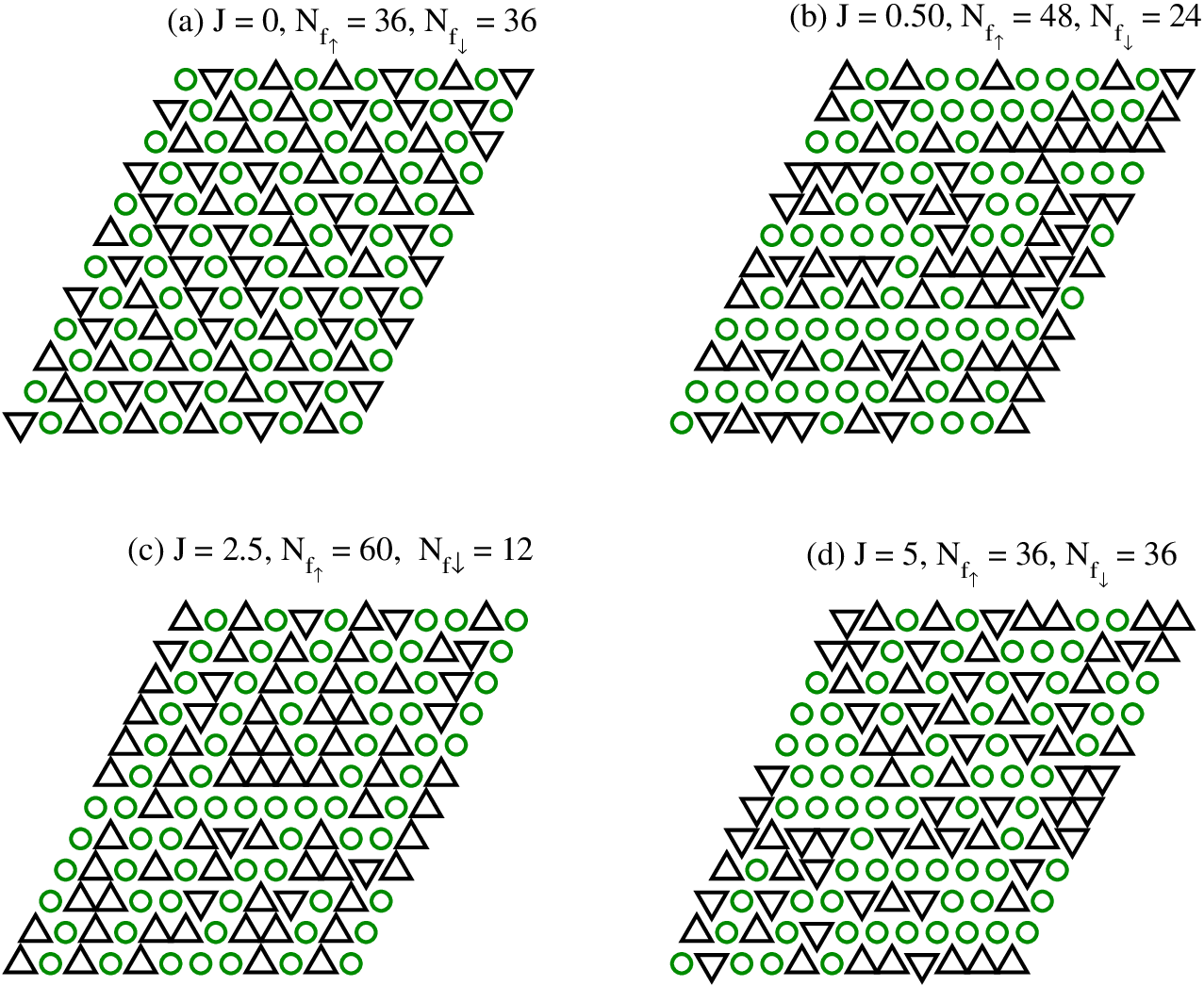}~~~~\includegraphics[trim=0.5mm 0.0mm 0.5mm 0.1mm,clip,width=8.9cm,height=6.0cm]{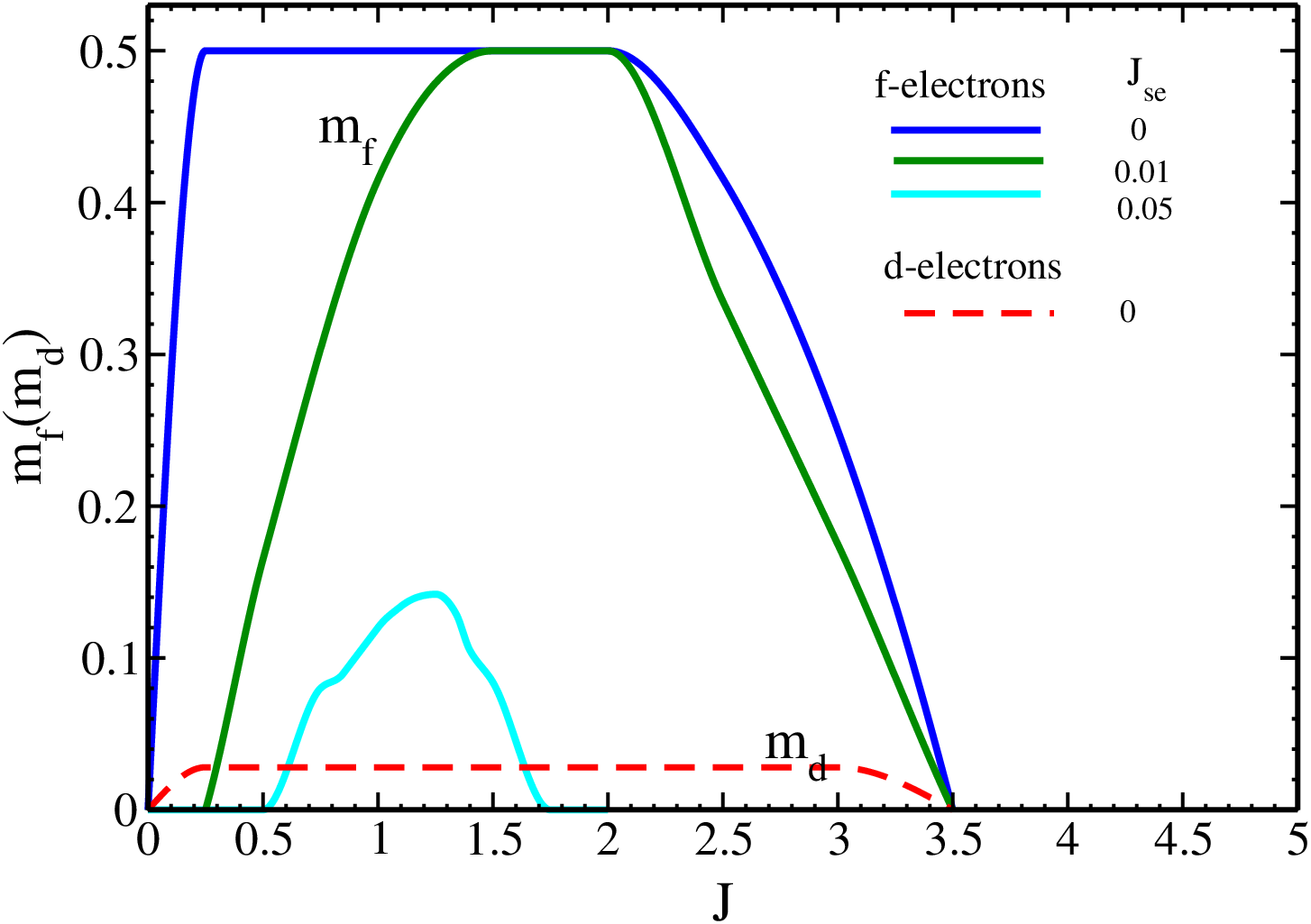}
{{\bf (i)}~~~~~~~~~~~~~~~~~~~~~~~~~~~~~~~~~~~~~~~~~~~~~~~~~~~~~~~~~~~~~~~~~~~~~~~~~~~~~~~~~~~~~~~~~~~~~~~~~~~~~~~~{\bf (ii)}}
\end{center}

\begin{center}
\includegraphics[trim=0.4mm 0.4mm 0.4mm 0.1mm,clip,width=8.9cm,height=4.0cm]{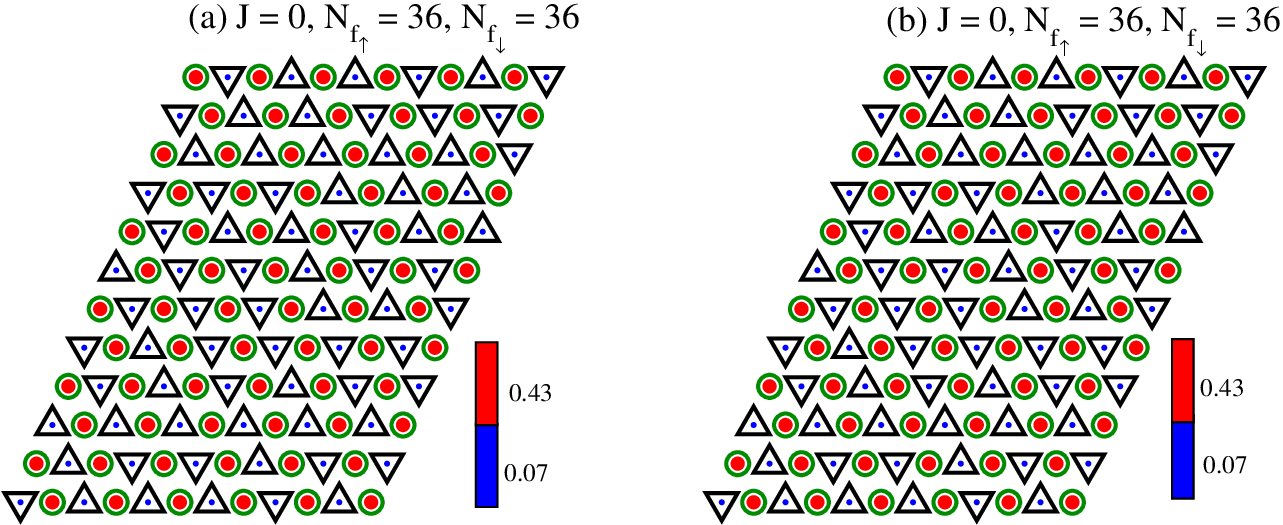}~~~~\includegraphics[trim=0.4mm 0.4mm 0.4mm 0.1mm,clip,width=8.9cm,height=4.0cm]{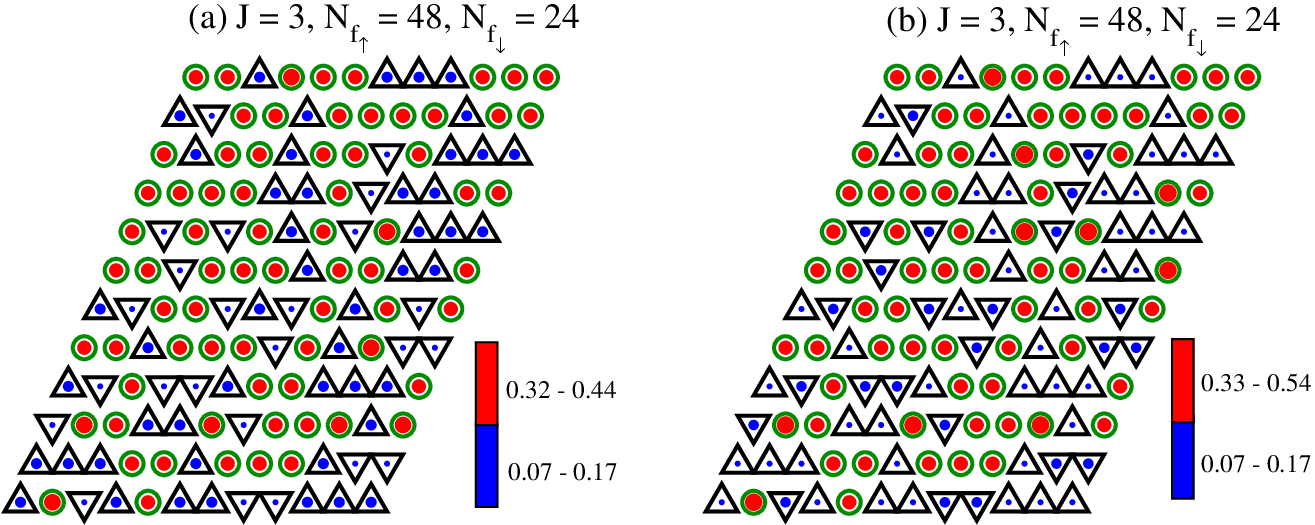}
{{\bf (iii)}~~~~~~~~~~~~~~~~~~~~~~~~~~~~~~~~~~~~~~~~~~~~~~~~~~~~~~~~~~~~~~~~~~~~~~~~~~~~~~~~~~~~~~~~~~~~~~~~~{\bf (iv)}}
\end{center}
}
\caption{(Color online) {\bf One-fourth filled case ($n_{f} + n_{d} = 1$) :} Triangle-up and triangle-down correspond to the sites occupied by up-spin and down-spin $f$- electrons respectively. Open green circles correspond to the unoccupied sites. The color coding and radii of the circles indicate the $d$-electron density profile. {\bf (i)} The ground-state magnetic configurations of $f$- electrons for $J_{se} = 0.01$, $n_{f} = \frac{1}{2}$, $n_{d} = \frac{1}{2}$, $U = 5$, $U_{f} = 10$ and for various values of $J$. {\bf (ii)} Variation of magnetic moment of $d$- and $f$- electrons with exchange correlation $J$ at different values of $J_{se}$ for $n_{f} = \frac{1}{2}$, $n_{d} = \frac{1}{2}$ at $U=5$, $U_{f} = 10$. Magnetic moment of $f$- and $d$- electrons are shown by solid and dash lines respectively. Note that ${m_{d}} = 0$ for finite values of $J_{se}$.} {\bf (iii)} (a) Up-spin and (b) down-spin $d$- electron densities are shown on each site for $n_{f} = \frac{1}{2}$, $n_{d} = \frac{1}{2}$, $U = 5$, $U_{f} = 10$ , $J = 0$ and $J_{se} = 0.01$. {\bf (iv)} (a) Up-spin and (b) down-spin $d$- electron densities are shown on each site for $n_{f} = \frac{1}{2}$, $n_{d} = \frac{1}{2}$, $U = 5$, $U_{f} = 10$ $J = 3$ and $J_{se} = 0.01$. 
\end{figure*}

We have studied the effect of $J_{se}$ on variation of magnetic moment of $d$- electrons ${m_{d}}$ $(= \frac{M_{d}}{N} = \frac{\large(N_{d_{\uparrow}}~ - ~N_{d_{\downarrow}}\large)}{N}\large)$ and magnetic moment of $f$- electrons ${m_{f}}$ $( = \frac{M_{f}}{N} =  \frac{\large(N_{f_{\uparrow}}~ - ~N_{f_{\downarrow}}\large)}{N}\large)$, at a fixed value of $U$, $U_{f}$ and $J$ for $n_{f} = 1~\large(n_{f} = \frac{N_{f}}{N} = \frac{\large(N_{f_{\uparrow}}~ + ~N_{f_{\downarrow}}\large)}{N}\large)$. We have also studied the density of $d$- electrons at each site for the above case. Fig.$1$(i) shows the variation of magnetic moment of $d$- and $f$- electrons with number of $d$- electrons $N_d$ for three different values of $J_{se}$ i.e. $J_{se} = 0$, $0.05$ and $0.1$ at a fixed value of $J = 5$, $U = 5$ and $U_{f} = 10$. We are studying the presence of magnetic phases for various values of $N_d$, starting from $N_d = 144$ to $N_d = 0$ at different values of $J_{se}$. We observed that for $N_d = 144$, it is AFM phase and remains AFM upto $N_d =N_{d}^{*}$ , below $N_{d}^{*}$ it is no more AFM and a net magnetic moment exists. Value of $N_{d}^{*}$ depends upon $J_{se}$. Variation of $N_{d}^{*}$ with $J_{se}$ for $U = 5 = J$, $U_f = 10$ and $n_{f} = 1$, shown in inset of Fig.$1$(i). We have noted that the ground state is Neel ordered anti-ferromagnetic (AFM) for $N_d = 144$, irrespective of the value of $J_{se}$.  As seen from Hamiltonian for $U = 5 = J$, there is no on-site coulomb repulsion between $d$- and $f$- electrons of the same spins, so up-spin $d$- electrons are more likely to occur at sites with up-spin $f$- electrons and same is true for down-spin $d$- electrons. Fig.$1$(i) also shows that the magnetic moment of $f$- electrons and $d$- electrons start increasing at value of $N_d$ below $70$ for $J_{se} = 0$, below $N_d = 65$ for $J_{se} = 0.05$ and below $N_d = 44$ for $J_{se} = 0.1$. It means larger the value of $J_{se}$, lower is the value of $N_d$ below which magnetic moment of $f$- and $d$- electrons start increasing. At $N_d = 144$, FM arrangement of $f$- electrons is not energetically favourable as $d$- electron's motion is prohibited by the Pauli exclusion principle whereas in an AFM arrangement system gains superexchange energy due to virtual hopping of $d$- electrons. As we lower $N_d$, sites with empty d-level appear and it becomes possible for $d$- electrons to move and gain kinetic energy. This kinetic energy gain then competes with super-exchange to decide the ground state magnetic ordering. As we lower $N_d$ from 144, hopping of $d$- electrons increases. However, the overall phase remains AFM (with zero magnetic moment) due to dominant superexchange interactions until $N_d$ reaches 65 (which we call $N_{d}^{\ast}$) (for $J_{se} = 0.05$). Below $N_{d}^{\ast}$, system develops a finite magnetic moment as parallel spin arrangement of $f$- electrons at neighbouring sites facilitates $d$- electron's hopping further. As we keep on decreasing $N_d$, a stage reaches where the system becomes totally FM because now band energy gain of $d$- electrons in the FM background totally overcomes the super-exchange energy gain of $f$- electrons. With further decrease of $N_d$ toward lower values of $N_d$( in this case $N_d < 12$) , system re-enters the AFM phase as the band energy gain of few $d$- electrons does not remain sufficient to overcome the energy gain due to super-exchange interactions of the AFM phase. Therefore the system prefers the AFM arrangement of $f$- electrons at low $N_d$ values which we call re-entrant AFM phase. 

          We have also studied the density of $d$- electrons at each lattice site. Fig.$1$(ii) shows the density of $d$- electrons at each site for the parameters taken in Fig.$1$(i) for $J_{se} = 0.1$ at $N_d = 40$. Density of $d$- electrons at each site strongly depends on the value of exchange correlation $J$ (between $d$- and $f$- electrons) and super-exchange interation $J_{se}$ between localized electrons. It is clear from Hamiltonian that for $U = 5 = J$, there is no onsite coulomb repulsion between $d$- and $f$- electrons of the same spins. Hence up-spin $d$- electrons density is more at sites having up-spin $f$- electrons. Effects of super-exchange interaction $J_{se}$, exchange interaction $J$, and on-site Coulomb repulsion $U$ on density of $d$- electrons can be understood in the following ways. It is seen from Fig.$1$(ii) that there are some sites having more $d$- electron densities than other sites. Super-exchange interaction $J_{se}$  arranges $f$- electrons in such a pattern that corresponds to minimum energy state, due to which each nearest-neighbour(NN) of a site ( each site have 6 NN) have opposite spin of $f$- electrons. If all the NN have spins opposite to concerned site then it becomes very difficult for $d$- electrons to hop from concerned site to NN sites, as there is finite onsite coulomb repulsion present between $d$- and $f$- electrons of the opposite spins. The random distribution of $f$- electrons and random values of $d$- electron densities is a consequence of the competition between superexchange interaction $J_{se}$, exchange interaction $J$ and on-site Coulomb repulsion $U$. We have also studied ground state magnetic phase diagram of up-spin and down-spin $f$- electrons, magnetic moments of $d$- and $f$- electrons and the density of $d$- electrons on each site for the range of values of parameters $J$, $U$, $U_{f}$ and $J_{se}$ for two cases (i) $n_{f} + n_{d} = 1$ (one-fourth filled case) and (ii) $n_{f} + n_{d} = 2$ (half filled case). We have chosen large value of $U_{f}$ so that double occupancy of $f$- electrons is avoided.

\subsection{One-fourth filled case ($n_{f} + n_{d} = 1$):}

In Fig.$2$(i) the ground state magnetic configurations of up-spin and down-spin $f$- electrons are shown (for one-fourth filled case) for $J_{se} = 0.01$, $U = 5$, $U_{f} = 10$ and for different $J$ values. Due to superexchange interaction between $f$- electrons, no ordered configurations of $f$- electrons are observed as explained already.

Fig.$2$(ii) shows the variation of magnetic moment of $d$- electrons ($m_{d}$) and magnetic moment of $f$- electrons ($m_{f}$) with exchange correlation $J$ at three different values of $J_{se}$ i.e $J_{se} = 0$, $0.01$, and $0.05$,  for $U = 5$ and $U_{f} = 10$. For $J = 0$, the on-site interaction energy between $d$- and $f$- electrons is same irrespective of their spins. Hence the ground state configuration is AFM type as possible hopping of $d$- electrons minimizes the energy of the system. It is clearly shown in the variation of $d$- electron density at each site in Fig.$2$(iii). For finite but small value of $J$, the on-site interaction energy between $d$- and $f$- electrons of same spins will be smaller in comparison to the on-site interaction energy between $d$- and $f$- electrons of opposite spins. So few sites with FM arrangement of spin-up $f$- electrons will be occupied by some down-spin $d$- electrons and some up-spin $d$- electrons. With this arrangement there is finite hopping possible for $d$- electrons which increases its kinetic energy and hence total energy of system goes down. Therefore the $m_d$ and $m_f$ increase with increasing $J$. It is also seen from Fig.$2$(ii) clearly that as value of $J_{se}$ increases, the magnetic moments of both $d$- and $f$- electrons decrease, as $J_{se}$ favours AFM arrangement of $f$- electrons and also favours AFM arrangement of $d$- electrons because on-site interaction between $d$- and $f$- electrons of same spins is  $(U - J)$ and on-site interaction between $d$- and $f$- electrons of opposite spins is $U$. There is no magnetic moment for $d$- electrons for $J_{se} = 0.01$ and $J_{se} =0.05$.

Figs.$2$(iii) and $2$(iv) show the density of $d$- electrons at a fixed value of $U = 5$, $U_{f} = 10$ and for  $J = 0$ and $J = 3$ respectively. When $J = 0$ the interaction between $d$- and $f$- electrons is same irrespective of their spins, so the density of $d$- electrons at sites occupied by f-electrons are same, while it is maximum at unoccupied sites. With the increase in $J$ value density of $d$- electrons at sites where $f$- electrons of same spin are present increases and at empty sites it decreases, because as $J$ increases, the interaction $(U-J)$ between $d$- and $f$- electrons of the same spins decreases. Also the random distribution of $d$- electron densities at some sites is, of course, due to the superexchange interaction between $f$- electrons.

\begin{figure*}
{
\begin{center}
\includegraphics[trim=0.5mm 0.5mm 0.5mm 0.0mm,clip,width=8.9cm,height=6.0cm]{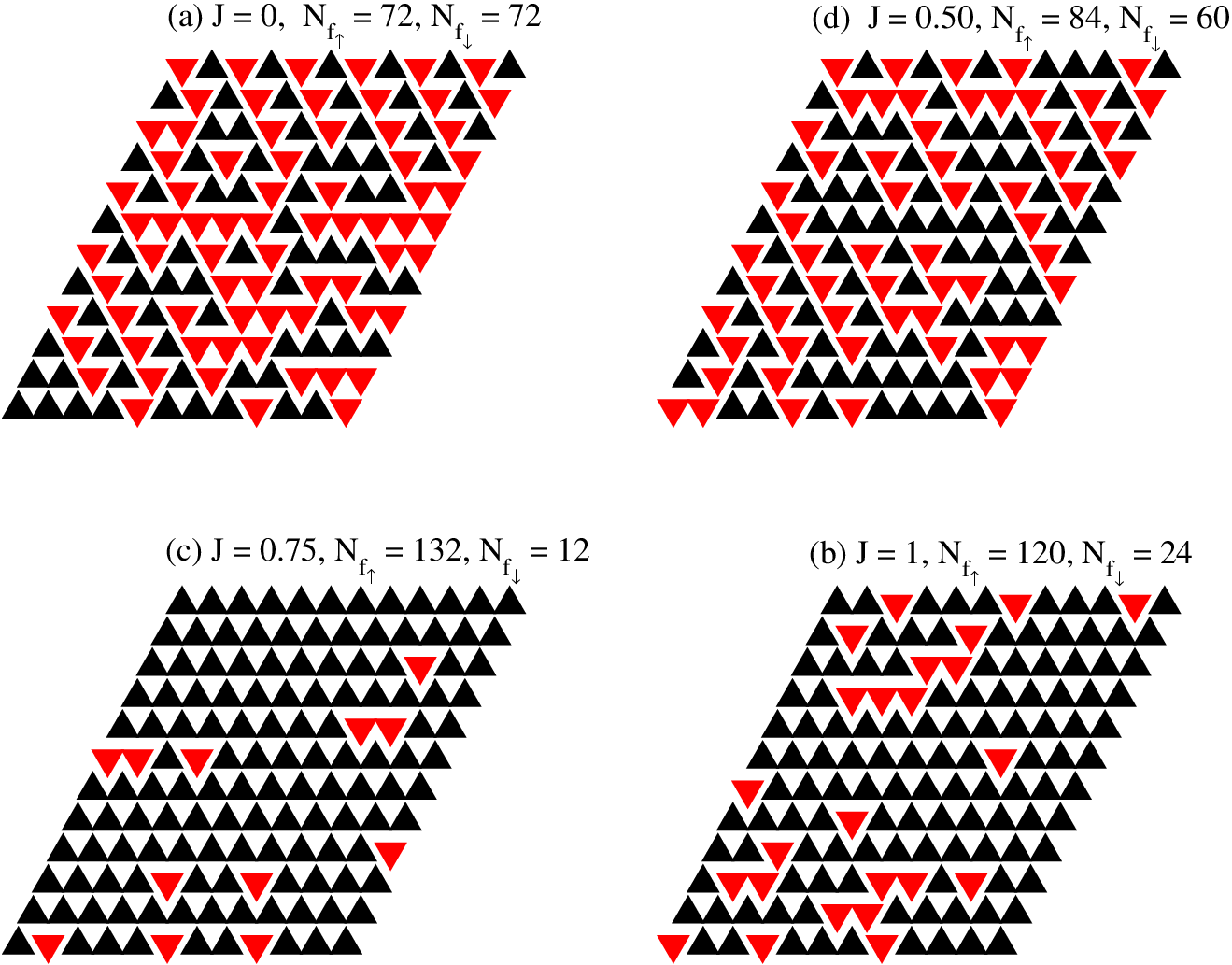}~~~~\includegraphics[trim=0.5mm 0.0mm 0.5mm 0.0mm,clip,width=8.9cm,height=6.0cm]{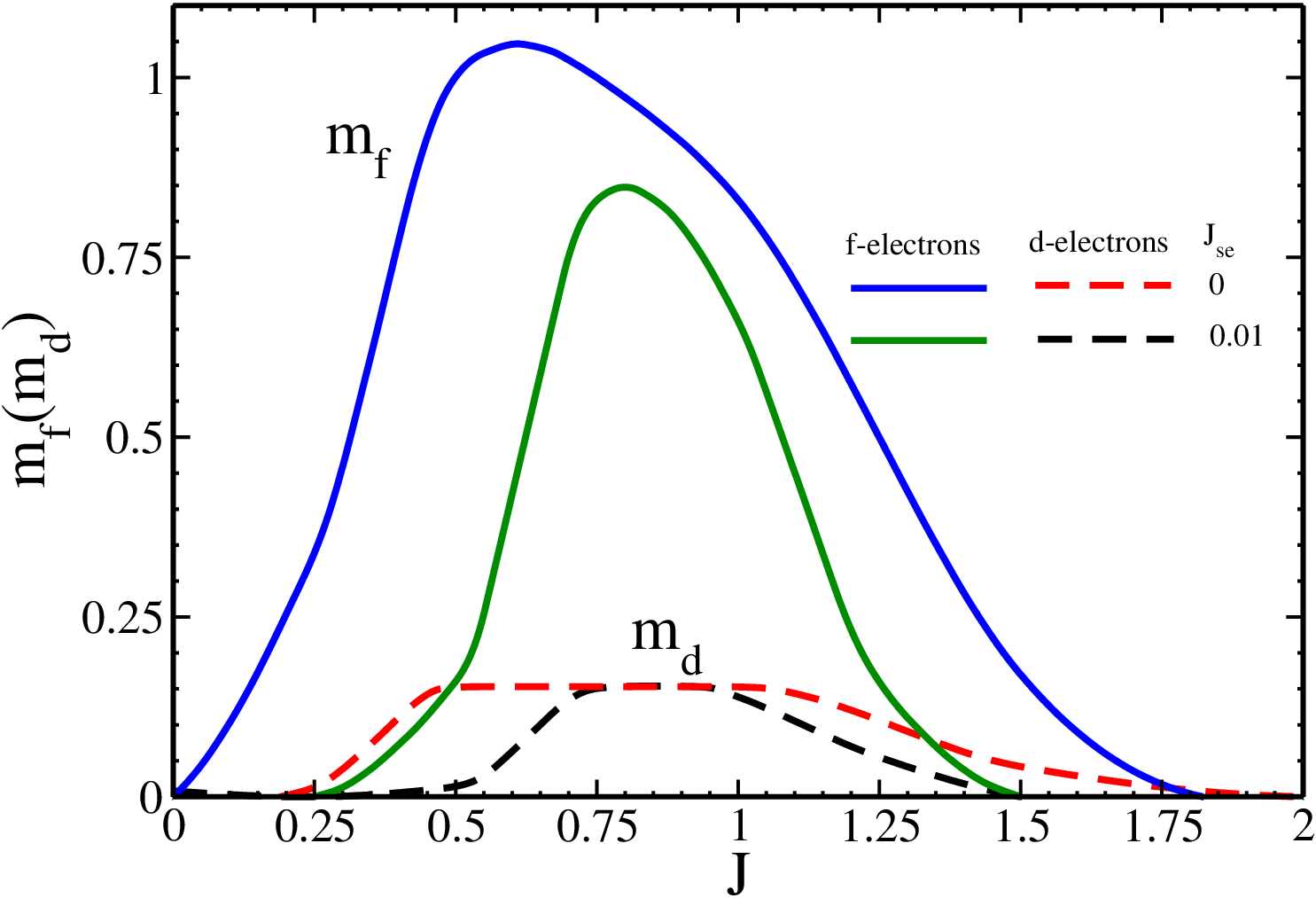}
{{\bf (i)}~~~~~~~~~~~~~~~~~~~~~~~~~~~~~~~~~~~~~~~~~~~~~~~~~~~~~~~~~~~~~~~~~~~~~~~~~~~~~~~~~~~~~~~~~~~~~~~~~~~~~~~~~~~{\bf (ii)}}
\end{center}

\begin{center}
\includegraphics[trim=0.4mm 0.4mm 0.4mm 0.1mm,clip,width=8.9cm,height=4.0cm]{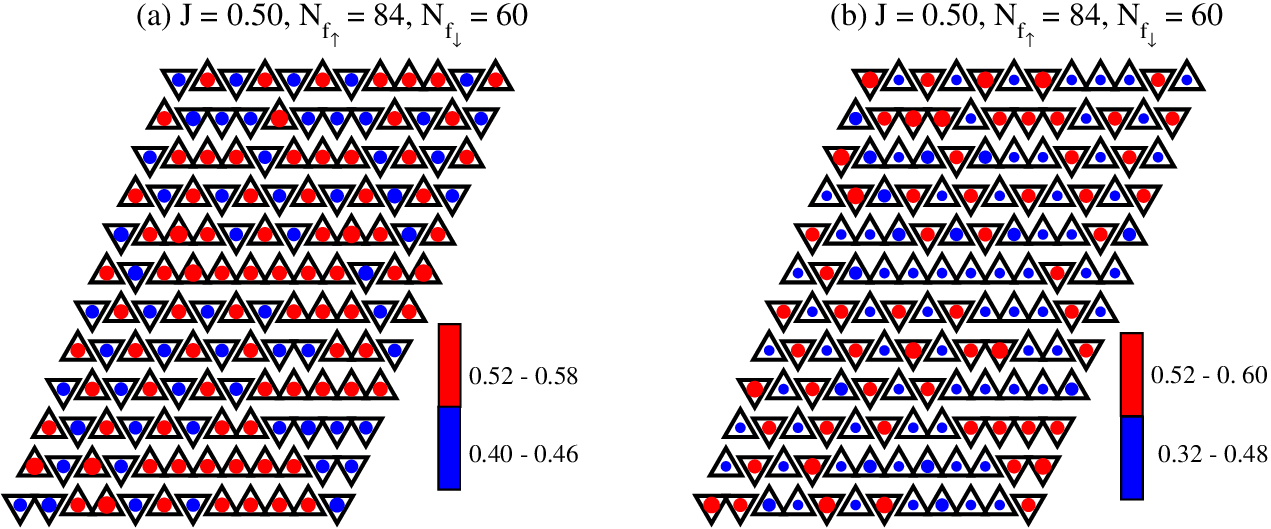}~~~~\includegraphics[trim=0.4mm 0.4mm 0.4mm 0.1mm,clip,width=8.9cm,height=4.0cm]{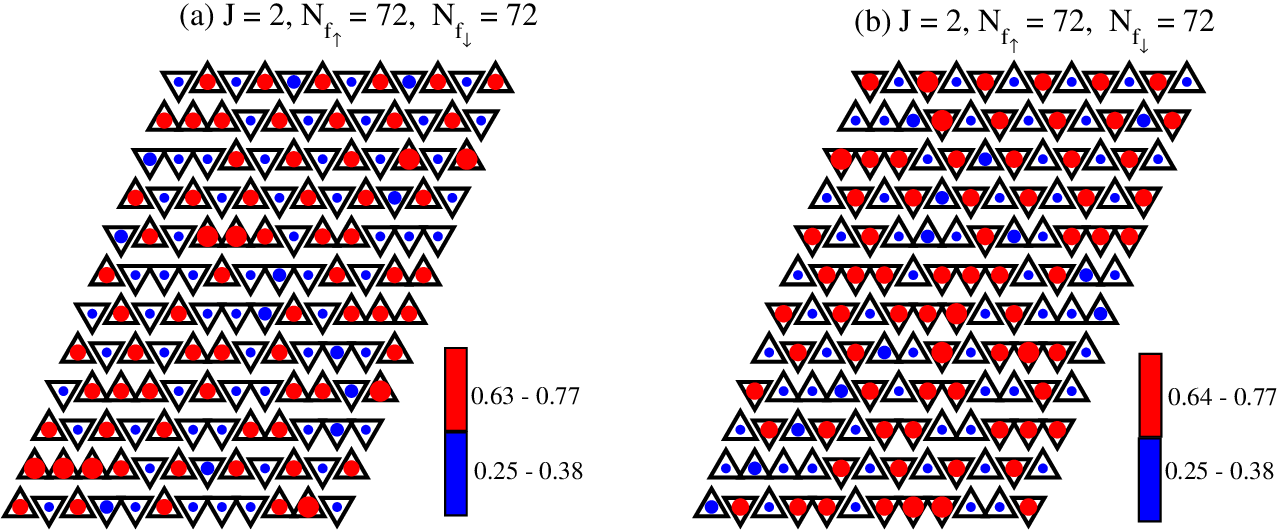}
{{\bf (iii)}~~~~~~~~~~~~~~~~~~~~~~~~~~~~~~~~~~~~~~~~~~~~~~~~~~~~~~~~~~~~~~~~~~~~~~~~~~~~~~~~~~~~~~~~~~~~~~~~~~~~~~~~~~~{\bf (iv)}}
\end{center}
}
\caption{(Color online) {\bf Half filled case ($n_{f} + n_{d} = 2$) :} Triangle-up and triangle-down correspond to the sites occupied by up-spin and down-spin $f$- electrons respectively. The color coding and radii of the circles indicate the $d$- electron density profile. {\bf (i)} The ground-state magnetic configurations of $f$- electrons for $J_{se} = 0.01$, $n_{f} = 1$, $n_{d} = 1$, $U = 5$, $U_{f} = 10$ and for various values of $J$. {\bf (ii)} Variation of magnetic moment of $d$- and $f$- electrons with exchange correlation $J$ at different values of $J_{se}$ for $n_{f} = 1$, $n_{d} = 1$ at $U=5$, $U_{f} = 10$. Magnetic moment of $f$- and $d$- electrons are shown by solid and dash lines respectively.(note that $m_d = 0$ for finite values of $J_{se}$. {\bf (iii)} (a) Up-spin and (b) down-spin $d$- electron densities are shown on each site for $n_{f} = 1$, $n_{d} = 1$, $U = 5$, $U_{f} = 10$ $J = 0.50$ and $J_{se} = 0.01$. 
{\bf (iv)} (a) Up-spin and (b) down-spin $d$- electron densities are shown on each site for $n_{f} = 1$, $n_{d} = 1$, $U = 5$, $U_{f} = 10$ $J = 2$ and $J_{se} = 0.01$.}
\end{figure*}

\subsection{Half-filled case ($n_{f}+n_{d}=2$):}

In Fig.$3$(i) the ground state magnetic configurations of up-spin and down-spin $f$- electrons are shown (for half-filled case) for $J_{se} = 0.01$, $U = 5$, $U_{f} = 10$ and for different $J$ values. Due to superexchange interaction between $f$- electrons, no ordered configurations of $f$- electrons are observed. Fig.$3$(ii) shows the variation of magnetic moment of $d$- electrons ($m_{d}$) and magnetic moment of $f$- electrons ($m_{f}$) with exchange correlation $J$ at two different values of $J_{se}$ i.e $J_{se} = 0$, and $0.01$, for $U=5$ and $U_{f} = 10$.  For $J = 0$, the on-site interaction energies between $d$- and $f$- electrons are same irrespective of their spins. Hence the ground state configuration is AFM type as possible hopping of $d$- electrons minimizes the energy of the system, which have been explained already.  For finite but small value of $J$, the on-site interaction energy between $d$- and $f$- electrons of same spins will be smaller in comparison to the on-site interaction energy between $d$- and $f$- electrons of opposite spins, so few sites with FM arrangement of spin-up $f$- electrons will be occupied by some down-up $d$- electrons and some up-spin $d$- electrons. With this arrangement there is finite hopping possible for $d$- electrons which increases its kinetic energy and hence total energy of system decreases. Therefore the $m_d$ and $m_f$ increase with increasing $J$. It is also seen from Fig.$2$(ii) (one-fourth filled case) clearly that as value of $J_{se}$ increases the magnetic moments of both $d$- and $f$- electrons decrease, as $J_{se}$ favour AFM arrangement of $f$- and $d$- electrons. For larger value of $J$, on-site interaction between $d$- and $f$- electrons with same spin decreases as coupling strength is $U - J$ (see Hamiltonain). So $d$- electrons would prefer sites having $f$- electrons with same spin. In this case, the AFM arrangement of $f$- electrons is favoured over FM as the system gains super-exchange energy due to the possibility of virtual hopping of $d$- electrons to neighbouring sites. Thus system re-enters AFM phase at higher values of $J$. Figs.$3$(iii) and $3$(iv) show the density of $d$- electrons at a fixed value of $U = 5$, $U_{f} = 10$ and for  $J = 0.5$ and $J = 2$ respectively. Fig.$3$(iii) shows that density of $d$- electrons with up-spin is slightly  more at sites with up-spins $f$- electrons as compared to sites with down-spin $f$- electrons. This is because of the fact that there is large value of on-site interaction energy ($U = 5$) between $d$- and $f$- electrons of the opposite spins as compared to on-site interaction energy($U - J = 4.50$) between $d$- and $f$- electrons of the same spins.

In conclusion, the ground state magnetic properties of two dimensional spin-1/2 FKM on a triangular lattice are studied for a range of parameter values like $d$- and $f$- electrons fillings, superexchange interaction $J_{se}$, on-site Coulomb correlation $U$ and exchange correlation $J$ etc. Extending the model to include the super-exchange interaction $J_{se}$ between $f$- electrons occupying nearest neighboring sites leads to interesting results: in particular, it strongly favours the $AFM$ arrangement of $f$- electrons and also plays a role in inducing AFM coupling between $d$- electrons.  We have found that the magnetic moments of $d$- and $f$- electrons depend strongly on the values of $J$ and on the number of $d$- electrons $N_d$. In half-filled case, a very small value of $J_{se}$ ($J_{se} > 0.01$) makes the system AFM and hence no net magnetic moment for $d$- or $f$- electrons is observed for $J_{se} > 0.01$ where as in one-fourth filling case a higher $J_{se}$ is needed to make the system AFM.  These results are quite relevant for study of triangular lattice systems mentioned above.

$Acknowledgments.$ SK acknowledges MHRD, India for a research fellowship. UKY acknowledges support form UGC, India  via Dr. DS Kothari Post-doctoral Fellowship scheme.

\end{document}